# Direct observation of nanometer-scale Joule and Peltier effects in phase change memory devices

Kyle L. Grosse[1,2], Feng Xiong[2,3,4], Sungduk Hong[2,3], William P. King[1,2,4,5,*], and Eric Pop[2,3,4,*]

[1]*Dept. of Mechanical Science & Eng., Univ. Illinois at Urbana-Champaign, IL 61801, USA*

[2]*Micro and Nanotechnology Lab, Univ. Illinois at Urbana-Champaign, IL 61801, USA*

[3]*Dept. of Electrical & Computer Eng., Univ. Illinois at Urbana-Champaign, IL 61801, USA*

[4]*Beckman Institute, Univ. Illinois at Urbana-Champaign, IL 61801, USA*

[5]*Dept. of Materials Science & Eng., Univ. Illinois at Urbana-Champaign, IL 61801, USA*

We measure power dissipation in phase change memory (PCM) devices by scanning Joule expansion microscopy (SJEM) with sub-50 nm spatial and 0.2 K temperature resolution. The temperature rise in the $Ge_2Sb_2Te_5$ (GST) is dominated by Joule heating, but at the GST-TiW contacts it is a combination of Peltier and current crowding effects. Comparison of SJEM and electrical characterization with simulations of the PCM devices uncovers a thermopower ~350 µV/K for 25 nm thick films of face centered-cubic crystallized GST, and contact resistance ~$2.0 \times 10^{-8}$ $\Omega \cdot m^2$. Knowledge of such nanoscale Joule, Peltier, and current crowding effects is essential for energy-efficient design of future PCM technology.

* wpk@illinois.edu; epop@illinois.edu

Phase change memory[1] (PCM) is a non-volatile memory technology with potential for fast (sub-nanosecond)[2] and low power (femtojoule)[3,4] operation. PCM has potential to replace dynamic random-access memory (DRAM) and Flash memory in future electronics.[5] Data in chalcogenide-based PCM, such as $Ge_2Sb_2Te_5$ (GST), are stored by the large ratio ($>10^3$) in electrical resistance between amorphous and crystalline states of the material. Reversible switching between phases is typically driven by Joule heating; however, Peltier,[6] Seebeck,[7] and Thomson[8] effects have been observed to contribute to phase change.[9] Previous studies have shown the thermopower for bulk and thin-film face-centered cubic (fcc) GST is large (200-400 µV/K).[7,10,11] Electrical contacts and thermal interfaces to GST are also important for heat generation and thermal confinement of GST devices.[12-15] Recent work has indirectly and separately measured the role of interfaces[12] and thermoelectric effects[6,8] in GST devices. These are essential, because electrical and thermal interfaces could reduce PCM programming power[12,13] by 20-30%, and thermoelectric effects may reduce power consumption[9] an additional 20-40% depending on the thermopower of thin GST films. However, direct observations of Joule, Peltier, and current crowding (interface) effects at the nanometer-scale in a PCM device are still lacking.

In this study, we measured the nanometer-scale temperature distributions in lateral PCM devices using scanning Joule expansion microscopy (SJEM),[16-19] which is an atomic force microscopy (AFM) based thermometry technique. The measurement's spatial and temperature resolutions were sub-50 nm and ~0.2 K. Joule heating dominated the temperature rise of the GST, while the temperature rise at the TiW contacts consisted of Joule, Peltier, and current crowding effects. Transfer length method (TLM) measurements on devices with varying lengths yielded GST electrical resistivity ($3.7 \pm 0.5 \times 10^{-4}$ Ω m) in the fcc phase, and GST-TiW electrical contact resistivity ($2.0 \pm 1.3 \times 10^{-8}$ $\Omega$ m$^2$). Comparing SJEM measurements to a finite element analysis (FEA) model, uncovered the thermopower ($350 \pm 150$ $\mu$V K$^{-1}$) of a 25 nm thick film of fcc-GST.

Figure 1(a) shows the lateral GST test devices. GST (25 nm) was deposited on 300 nm thermal SiO$_2$ with a highly p-doped Si substrate. For electrical contact, 100 nm TiW (10/90 % weight) was patterned by photolithography and deposited by sputtering. Devices were encapsulated by electron-beam evaporation of 10 nm SiO$_2$. Fabrication was completed by spin coating 60 nm of polymethyl methacrylate (PMMA) on the samples. The PMMA protects the devices from oxidation, and amplifies thermo-mechanical expansions of the PCM device during operation, as needed for the SJEM technique.[18] Before starting the SJEM measurements we crystallized the GST into the fcc phase by baking the entire device on a hot plate at 200 °C for 5 minutes.[7,20,21] X-ray diffraction (XRD) analysis confirms the fcc films (see supplement[22]) albeit with relatively smaller grain size (~8 nm) than a previous study (~15-20 nm),[23] and with measured electrical resistivity in the generally accepted range for thin fcc films ($10^{-4} - 10^{-3}$ $\Omega$ m).[1,3,24]

Figure 1(a) also shows a schematic of the SJEM experiment. A sinusoidal waveform at 28 kHz with peak voltage $V$ was applied to resistively heat the device. The associated thermo-mechanical expansions of the sample were measured by the AFM cantilever, laser, and photodiode. A lock-in amplifier at the heating frequency $f_H$ with a low-pass filter bandwidth of 3-27 Hz recorded the peak surface expansion $\Delta h$. The spatial resolution was ~50 nm and temperature resolution was ~0.2 K based on previous reports[16,19] and similar sample geometries with.[18] The spatial resolution is further discussed in the supplement.[22] SJEM can directly resolve current crowding and Peltier effects due to current flow between the GST and TiW as the current transfer length $L_T = 1.2 \pm 0.5$ $\mu$m, calculated below, between the GST-TiW is greater than the spatial resolution.

Figure 1(b) shows the surface expansion $\Delta h$ overlaid on the topography of a device. The GST peak temperature rise $\Delta T$ is proportional to $\Delta h$ and is related by FEA modeling.[17,18] The device was biased with $V = \pm 8.9$ V at 28 kHz, and $\Delta h$ was recorded at $f_H = 56$ kHz, as Joule heating occurs at twice the applied frequency for a bipolar sine wave. Further increasing $f_H$ decreases $\Delta h$ as the thermal diffusion length decreases, which decreases the amount of material which thermo-mechanically expands.[19] At the micrometer-scale $\Delta h$ is uniform across the device indicating uniform lateral heating, electric field, and resistivity distribution. Figure 1(b) also reveals some rough TiW contact edges from the lift-off process, which were avoided during measurements to limit measurement artifacts.

Before analyzing the SJEM measurements, we obtained device and contact resistance information. Figure 2 shows TLM measurements of the device resistance $R_{DS}$ of 55 GST devices with lengths $L = 1.5$ to 10 μm and fixed width $W = 245$ μm (see Fig. 2 inset). The slope and $y$-axis intercept of a linear fit to measurements yields the sheet resistance $R_\square = 15 \pm 2$ kΩ/□ and twice the GST-TiW contact resistance per width $2R_C W = 35 \pm 11$ kΩ μm. Therefore, the GST resistivity $\rho_{GST} = 3.7 \pm 0.5 \times 10^{-4}$ Ω m for the fcc phase, similar to previous studies.[24,25] The GST-TiW electrical interface resistivity $\rho_C$ and current transfer length $L_T$ are calculated from:[26]

$$R_C \cdot W = \left( \rho_C / L_T \right) \coth(L_C / L_T) \tag{1}$$

$$L_T = \sqrt{\rho_C / R_\square} \quad . \tag{2}$$

Equations 1 and 2 yield $\rho_C = 2.0 \pm 1.3 \times 10^{-8}$ Ω m² and $L_T = 1.2 \pm 0.5$ μm, where the TiW contact length was $L_C >> L_T$. The contact resistance is larger than in a previous study for TiW with fcc GST[27] (~$10^{-9}$ Ω m²), which is attributed to the relatively lower quality of the sputtered TiW in this work,[22] and lower thermal budget of our devices. Additional information on the TLM method is also provided in the supplement.[22]

The thermo-mechanical expansion $\Delta h$ and corresponding temperature rise $\Delta T$ were predicted from a two-dimensional (2D) FEA model of the devices, used to interpret the SJEM measurements. A 2D model is appropriate as $W \gg L$ for our devices. The heat diffusion and Poisson equations were modified to account for thermoelectric transport[28,29] and were coupled with a thermo-mechanical model. The Fourier transform of the equations yielded the frequency response of the predicted $\Delta h$ and $\Delta T$. Additional information is given in the supplement.[22]

Figure 3(a) shows the measured and predicted $\Delta h$ for a device with $L = 1.5$ µm. The supplement[22] describes measurements for a device with $L = 7$ µm. A unipolar sine wave directed current flow (holes in p-type GST) across the device for $V = \pm 1.6$, $\pm 2.4$, and $\pm 3.2$ V. Measurements are an average of 32 scans with deviation smaller than the markers.

We observe heat generation at the GST-TiW interface due to current crowding and Peltier effects. Current crowding is independent of carrier flow direction and occurs at the GST-TiW interface due to a finite interface resistivity, $\rho_C$, between the GST channel and TiW contacts. On the other hand, the Peltier effect[30] is dependent on the direction of current flow through the GST-TiW junction and heats (cools) the junction as carriers flow into (out of) the contact due to the difference in GST and TiW thermopower. The TiW-GST interface properties were found by fitting the predicted and measured $\Delta h$. A $\rho_{GST} = 1.7 \times 10^{-4}$ Ω m and $\rho_C = 3 \times 10^{-9}$ Ω m$^2$ predicts $\Delta h$ and $R_{DS}$ which match measurements. Similar values were obtained for the device in the supplement.[22] Both $\rho_C$ and $\rho_{GST}$ are close to values obtained from TLM measurements summarized in

Fig. 2. We estimated the thermopower of GST in the fcc phase $S_{GST} \approx 350 \pm 150$ µV K$^{-1}$, by comparing measurements of the Peltier effect at the contacts in Fig. 3 with the FEA model.

Figure 3(b) shows the predicted $\Delta T$ and reveals the roles of Joule, Peltier, and current crowding effects of the PCM test device. Joule heating dominates power dissipation in the GST as expected, showing $\Delta T$ which scales with $V^2$. The majority of heat generation at the contacts is due to the finite $\rho_C$ and associated current crowding effect. A small temperature "spike" occurs at $|x| = 0.75$ µm for hole flow from the GST into the contact. The small hot spot forms due to heat generation at the contact, from current crowding and/or Peltier heating, combined with the low thermal conductance of the surrounding materials. Peltier heating and cooling is observed as the change in $\Delta T$ with hole flow direction. At $|V| = 1.6$ and 3.2 V the difference in $\Delta T$ with carrier flow at the channel edge is ~1.5 and 3 K (~63 and ~32 % of the channel $\Delta T$). The effect of different material properties on current crowding, Peltier effects, and hot-spot formation is further explored in the supplement.[22] The temperature resolution of 0.2-0.4 K, increasing with increasing bias, was determined by the predicted uncertainty in $\Delta T$ from the deviation of the measured $\Delta h$.

Figure 4 shows the fitting of predictions to measurements to determine the GST thermopower, $S_{GST}$. The difference in $\Delta h$ for hole flow left and right (due to Peltier heating and cooling of the contacts) is denoted $\Delta h_{Peltier} = \Delta h(j+) - \Delta h(j-)$, where $j+$ and $j-$ denote hole flow left and right. Figure 4(a) shows the measured and predicted $\Delta h_{Peltier}$ at the GST-TiW interface for $|V| = 1.6$, 2.4, and 3.2 V. In Fig. 4(a) the GST thermopower $S_{GST} = 250$, 500, and 500 µV K$^{-1}$ for $|V| = 1.6$, 2.4, and 3.2 V, corresponding to the best fits from Fig. 4(b). Figure 4(b) shows the coefficient of determination $R^2$ for predictions at both contacts for each bias condition. The average $R^2$ curve has a maximum $R^2 = 0.65$ which predicts $S_{GST} \approx 350 \pm 150$ µV K$^{-1}$ for fcc phase

GST, similar to previous studies.[7,10,11] The uncertainty in $S_{GST}$ was estimated from a 0.1 decrease below the maximum $R^2$.

In conclusion, we directly observed Joule, Peltier, and current crowding effects in a fcc-GST device using SJEM with ~50 nm spatial and ~0.2 K temperature resolution. The sheet and contact resistance of GST and GST-TiW were measured by TLM, and also confirmed by FEA simulation fitting against the SJEM data. Joule heating dominated power dissipation in the GST channel, while power dissipation at the GST-TiW contacts was a combination of Peltier and current crowding effects. Comparing measurements and modeling predictions, we obtained $S_{GST} \approx$ $350 \pm 150$ µV K$^{-1}$ for a 25 nm thick film of fcc-GST. The large measured thermopower of GST could reduce the energy consumption by >50 % in highly scaled PCM devices due to Peltier heating, compared to scenarios which only utilize Joule heating.[9] PCM energy consumption could be further reduced by optimizing the GST-metal contact[13] and the thermopower difference at these interfaces.[6] Such knowledge of nanometer-scale Joule, thermoelectric, and interface effects in GST devices should enable improvements in energy efficient designs of future PCM technology.

We gratefully acknowledge the help of S. Ozerinc, F. Lian, and M. Sardela. This work was in part supported by the National Science Foundation (NSF) grant ECCS 10-02026, and by the Materials Structures and Devices (MSD) Focus Center, under the Focus Center Research Program (FCRP), a Semiconductor Research Corporation entity. S.H. acknowledges support from a Samsung Fellowship.

**Figure Captions**

FIG. 1. (a) Schematic of phase change memory (PCM) device and scanning Joule expansion microscopy (SJEM) experiment. Lateral test devices consisted of 60 nm PMMA, 10 nm $SiO_2$, 100 nm TiW, 25 nm GST, and 300 nm $SiO_2$ on a Si substrate, from top to bottom. GST was crystallized into the face-centered cubic (fcc) phase by baking at 200 $^o$C for 5 min. SJEM operates by supplying a periodic voltage waveform to resistively heat the GST device, while atomic force microscopy (AFM) measures the resulting peak surface thermo-mechanical expansion $\Delta h$. (b) Measured $\Delta h$ overlaid on topography for one of the measured devices. The peak GST temperature rise $\Delta T$ is proportional to $\Delta h$ (also see refs. 17,18).

FIG. 2. Transfer length method (TLM) measurements of 55 GST devices. A linear fit to two-terminal measurements of device resistance $R_{DS}$ yields the GST sheet resistance $R_{\square}$ = 15 ± 2 k$\Omega$/$\square$ and twice the GST-TiW contact resistance $2R_C W$ = 35 ± 11 k$\Omega$ $\mu$m. The linear fit and standard deviation are shown by the solid red and dash-dot blue lines. The inset shows a top-view of the device geometry with GST channel length $L$ = 1.5 to 10 $\mu$m and a fixed width $W$ = 245 $\mu$m ($W \gg L$).

FIG. 3. (a) Measured and predicted $\Delta h$ and (b) predicted $\Delta T$ for $V$ = ±1.6, ±2.4, and ±3.2 V for the $L$ = 1.5 $\mu$m device. The edges of the GST-TiW contacts are marked by black vertical dashed lines. (a) Symbols show measured $\Delta h$ and lines show the predicted $\Delta h$. Current (hole) flow left and right are shown in red circles and solid lines, and blue triangles and dashed lines, respectively. The arrows indicate the hole flow direction with color and every second measurement is shown for clarity. (b) Predicted GST temperature rise $\Delta T$ for hole flow to the left (red solid line) and right (blue dashed line) is due to Peltier effects at the contacts. The black dashed lines and arrows are similar to (a).

FIG. 4. (a) Measured and predicted $\Delta h_{Peltier}$ at the GST-TiW interface for the $L = 1.5$ µm device. Three bias conditions are shown $|V| = 1.6$, 2.4, and 3.2 V in green dash-dot line and crosses; blue dashed line and triangles; and red solid line and dots. The GST thermopower $S_{GST} = 250$, 500, and 500 µV K$^{-1}$ for $|V| = 1.6$, 2.4, and 3.2 V. Bars located in the top-left show the standard deviation of the measurements over 32 scans. Measurements are shown by markers and predictions are shown as solid lines. (b) The coefficient of determination $R^2$ for predictions from both contacts. The three bias conditions are shown similar to (a). The average $R^2$ is shown as a black solid line. Negative $R^2$ indicates the measurement average is a better fit than predictions.

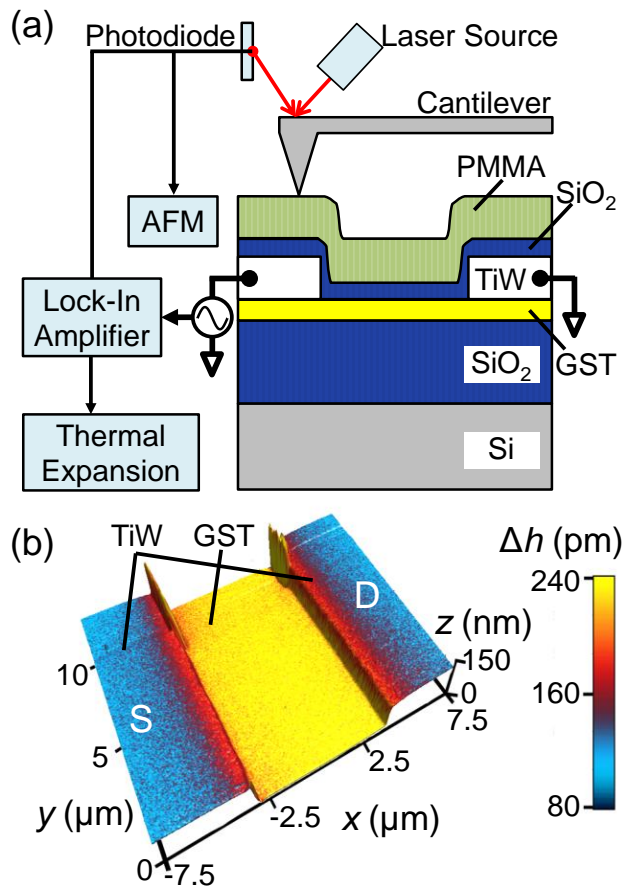

**(a)** Photodiode | Laser Source

Cantilever

PMMA | SiO₂

AFM

TiW

Lock-In Amplifier

SiO₂ | GST

Thermal Expansion

Si

**(b)** TiW | GST | Δh (pm)

D

z (nm)

150

S

0

7.5

2.5

x (μm)

y (μm)

0 -7.5

-2.5

240

160

80

**Figure 1**

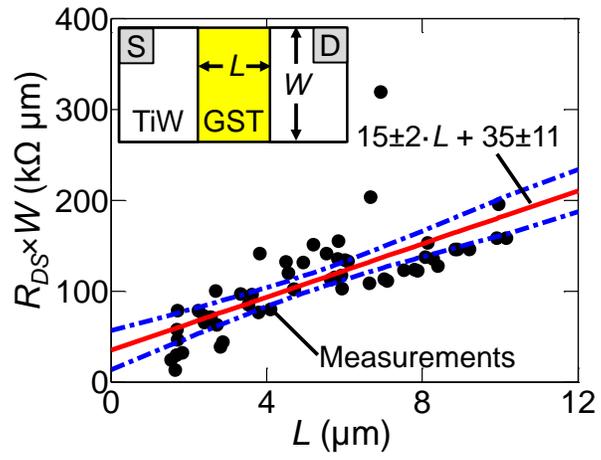

**Figure 2**

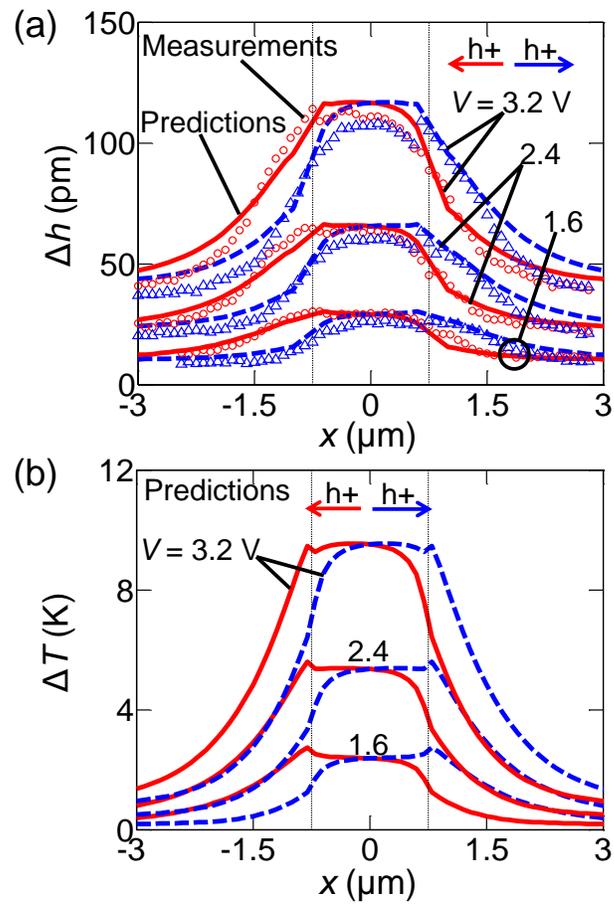

**Figure 3**

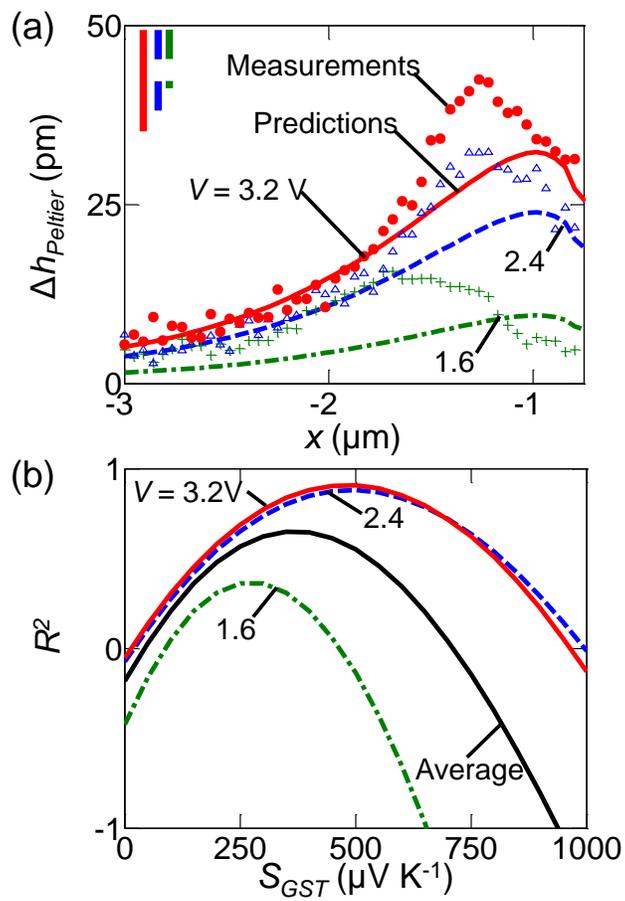

**Figure 4**

## Supplementary Materials for

by K.L. Grosse, F. Xiong, S. Hong, W.P. King, and E. Pop
University of Illinois at Urbana-Champaign, Urbana, IL 61801, U.S.A.

## 1. Transfer Length Method

Device resistance $R_{DS}$ was measured with two-terminal measurements, where the subscripts "D" and "S" refer to the two device contacts (drain and source). The channel length $L$ was measured using a combination of atomic force microscopy (AFM) and optical microscopy. The measured resistance $R$ is a series combination $R_{DS} + R_{Series}$, the latter including the leads, TiW contact pads, and TiW-probe contact. As shown in Table SI, the TiW resistivity is large and therefore contributes significantly to $R_{Series}$. Contacting the probes several times to the same TiW pad yielded $R_{Series} = 60 \pm 13 \ \Omega$, with $R_{Series} > 80 \ \Omega$ for a few mea-surements. Variation in device probe location adjusts $R_{Series}$ and $R$, contributing to the scatter seen in TLM measurements (Fig. 2 main text). The average $R_{Series}$ was subtracted from the measured resistance $R$ to yield $R_{DS}$ for the TLM measurements. An optical microscope (3,500x magnification) measured $L$ for each device at different points across the channel width. Four devices were measured by both AFM and the optical microscope. AFM measurements were within ~100 nm of optical measurements. The optical measurements showed $L$ deviated ~200 nm across the channel width.

## 2. Additional Device ($L$ = 7 μm)

Figure S1 shows the measurements and finite element analysis (FEA) predictions for a $L$ = 7 μm device. Figure S1(a) shows the measured and predicted peak surface expansion $\Delta h$, and Fig. S1(b) shows the device peak temperature rise $\Delta T$. The analysis is similar to the $L = 1.5$ μm device of the main text and duplicate details are omitted here. The device was resistively heated by applying an unipolar sine wave with $V = \pm 3.6$, $\pm 6.3$, and $\pm 8.9$ V at 28 kHz. Measurements are an average of 64 scans with standard deviations smaller than the symbol size for $|V| = 3.6$ and 6.3 V, and approximate twice the symbol size for $|V| = 8.9$ V. For the device in Fig. S1, $\rho_{GST} = 2.0 \times 10^{-4}$ Ω m, and $S_{GST} = 400 \pm 150$ μV K$^{-1}$. We attribute the dissimilar measured $\Delta h$ at the left and right contacts to dissimilar $\rho_C$ and heat dissipation at the contacts, likely due to non-uniform adherence between the TiW and GST. We found $\rho_C = 3$ and $13 \times 10^{-9}$ Ω m$^2$ for the right and left contacts, respectively. We are unable to discern if other devices exhibited non-uniform $\rho_C$ as the TLM measurement is an average of the 55 tested devices.

Joule heating and Peltier effects are evident in Fig. S1(b). Peltier heating and cooling is observed as the change in $\Delta T$ with hole flow direction. At $|V| = 3.6$ and 8.9 V the difference in $\Delta T$ with carrier flow at the GST-TiW edge is ~2.1 and ~5.6 K, respectively (~60% and ~26 % of the channel $\Delta T$). The temperature resolution of 0.2–0.7 K, increasing with increasing bias, was determined by the predicted uncertainty in $\Delta T$ from the deviation of the measured $\Delta h$.

Figure S2 shows the fitting of the predicted GST thermopower $S_{GST}$ to measurements. The average $R^2$ curve has a maximum $R^2 = 0.82$ which predicts $S_{GST} = 400 \pm 150$ μV K$^{-1}$. A similar value for $S_{GST}$ is predicted using the mean absolute percent error (MAPE) between measurements and predictions. The difference in $S_{GST}$ between the $L = 1.5$ and 7 μm devices is due to uncertainty in the measurement.

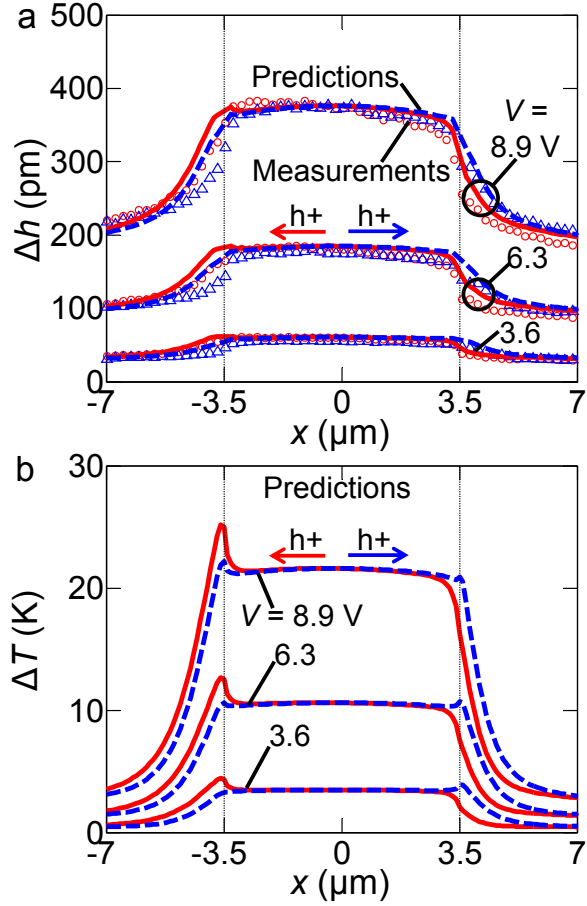

FIG. S1. (a) Measured and predicted $\Delta h$ and (b) predicted $\Delta T$ for $V = \pm 3.6$, $\pm 6.3$, and $\pm 8.9$ V for the $L = 7$ μm device. Black vertical dashed lines mark the edges of the GST-TiW contacts. (a) Symbols show the measured $\Delta h$ and lines show the predicted $\Delta h$. Current (hole) flow to the left and right is shown in red circles and solid lines, and blue triangles and dashed lines, respectively. Arrows indicate the hole flow direction with color and every fifth measurement is shown for clarity. (b) Predicted GST temperature rise $\Delta T$ for hole flow to the left (red solid line) and right (blue dashed line). Black dashed lines and arrows are similar to (a). The higher $\rho_C$ of the left contact for this device (13 vs. $3 \times 10^{-9}$ Ω m² for the right contact, see discussion in text) causes a hot spot to form at the left GST-TiW interface due to increased local heat generation.

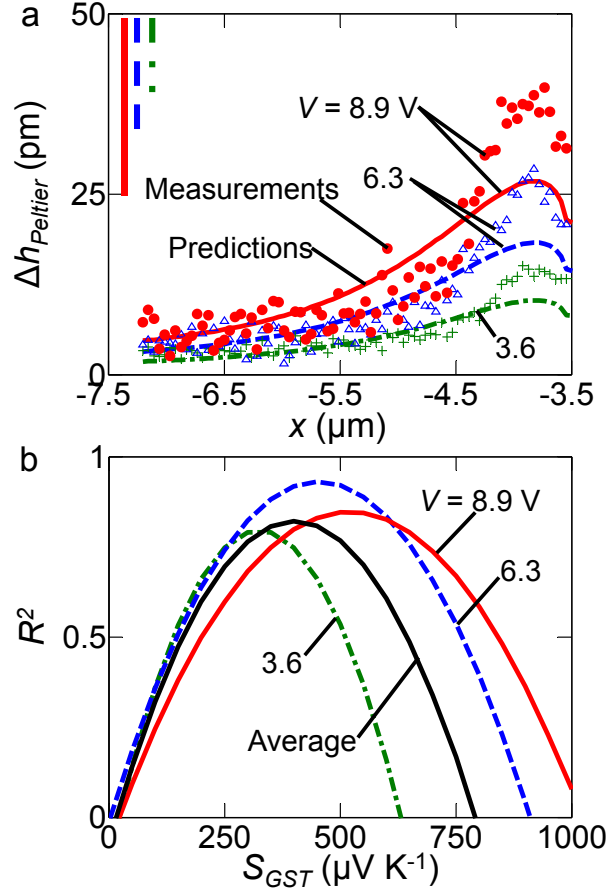

FIG. S2. (a) Measured and predicted $\Delta h_{Peltier}$ at the GST-TiW interface of the left contact, with $S_{GST} = 400$ µV K$^{-1}$ for the $L = 7$ µm device. Three bias magnitudes are shown $|V| = 3.6$, 6.3, and 8.9 V in green dash-dot line and crosses; blue dashed line and triangles; and red solid line and dots. Measurements are shown by markers and predictions are shown as solid lines. Bars in the top-left are the standard deviations of the measurement averaged over 64 scans. The Peltier effect causes the observed difference in heat generation with carrier flow. (b) The coefficient of determination $R^2$ for predictions from both contacts, with three bias conditions similar to (a). The average $R^2$ is a black solid line.

### 3. Model Equations

A two dimensional (2D) frequency domain thermoelectric-mechanical finite element analysis (FEA) model was developed to predict GST device behavior. The modified heat diffusion and Poisson equations are shown in Eqns. SE1 and SE2.[S1,S2]

$$\rho_d c_P \frac{\partial T}{\partial t} = \nabla(k + \sigma S^2 T)\nabla T + \nabla(\sigma S T)\nabla \phi + \sigma(S\nabla T\nabla\phi + [\nabla\phi]^2) \qquad \text{(SE1)}$$

$$\nabla\sigma(S\nabla T + \nabla\phi) = 0 \qquad \text{(SE2)}$$

The density, heat capacity, thermal conductivity, electrical conductivity, thermopower, temperature, and voltage are given by $\rho_d$, $c_P$, $k$, $\sigma$, $S$, $T$, and $\phi$. Equations SE3 and SE4 show the expected voltage and thermal waveforms.

$$\phi = \phi_0 + \phi_1 \cos(2\pi f t) \qquad \text{(SE3)}$$

$$T = T_0 + T_1 \cos(2\pi f t) + T_2 \cos(2\pi 2 f t) \qquad \text{(SE4)}$$

The subscripts denote the amplitude of $\phi$ and $T$ at the zero, first, and second harmonics. The frequency of the applied bias $f = 28$ kHz for all experiments. The predicted peak voltage applied to the device $V_{DS} = \text{sign}(\phi_0)(|\phi_0|+|\phi_1|)$ is applied to the non-grounded TiW edge in the model. The peak temperature rise of the GST $\Delta T = 2|T_1|$. The Fourier transform of Eqns. SE1 and SE2 with Eqns. SE3 and SE4 yields Eqn. SE5.

$$\nabla \begin{bmatrix} 4(k + \sigma S^2 T_0) & 2\sigma S^2 T_1 & 2\sigma S^2 T_2 & 4\sigma S T_0 & 2\sigma S T_1 \\ 2\sigma S^2 T_1 & 2k + \sigma S^2(2T_0 + T_2) & \sigma S^2 T_1 & 2\sigma S T_1 & \sigma S(2T_0 + T_2) \\ 2\sigma S^2 T_2 & \sigma S^2 T_1 & 2(k + \sigma S^2 T_0) & 2\sigma S T_2 & \sigma S T_1 \\ \sigma S & 0 & 0 & \sigma & 0 \\ 0 & \sigma S & 0 & 0 & \sigma \end{bmatrix} \nabla \begin{bmatrix} T_0 \\ T_1 \\ T_2 \\ \phi_0 \\ \phi_1 \end{bmatrix} =$$

$$\dots \begin{bmatrix} 2\sigma\left\{S(2\nabla T_0\nabla\phi_0 + \nabla T_1\nabla\phi_1) + 2(\nabla\phi_0)^2 + (\nabla\phi_1)^2\right\} \\ \sigma\left\{S(2\nabla T_0\nabla\phi_1 + 2\nabla T_1\nabla\phi_0 + \nabla T_2\nabla\phi_1) + 4\nabla\phi_1\nabla\phi_0\right\} - i4\pi\rho_d c_P T_1 f \\ \sigma\left\{S(\nabla T_1\nabla\phi_1 + 2\nabla T_2\nabla\phi_0) + (\nabla\phi_1)^2\right\} - i4\pi\rho_d c_P T_2 2 f \\ 0 \\ 0 \end{bmatrix} \qquad \text{(SE5)}$$

Equation SE5 was coded into the PDE physics and coupled with thermo-mechanical physics in COMSOL. Transforming and solving the equations in the frequency domain reduced computation time and convergence issues. Interface resistance and dissipation were implemented in COMSOL using home-built code.

## 4. Voltage Divider

The peak measured voltage $V$ and predicted device voltage $V_{DS}$ are related by:

$$V_{DS} = (V \times R_{DS}) / R \qquad (SE6)$$

The measured resistance $R = R_{Series} + R_{DS}$. The simulation predicts the device resistance $R_{DS}$ for the modeled devices based on $\rho_{GST}$, $\rho_C$, and $L$. Varying $\rho_{GST}$ and $\rho_C$ changes the predicted $\Delta h$ as the predicted device power dissipation changes. The best fit between measured and predicted $\Delta h$ determined $\rho_{GST}$ and $\rho_C$ which also predicted $R_{DS}$ and $V_{DS}$. For $L = 7$ and 1.5 μm devices, $R = 433$ and 198 Ω, $R_{DS} = 550$ and 279 Ω. The predicted $R_{DS}$ is larger than $R$, as the model simulates the probes are 200 μm from the GST channel, when actual probes were within ~100-150 μm of the channel. The overestimation of the TiW contact length adds ~50-100 Ω to $R_{DS}$.

## 5. Model Geometry

Figure S3 shows a schematic of the FEA model with boundary conditions. All surfaces were electrically insulated, thermally insulated, or mechanically constrained except those discussed below. The two TiW edges were fixed at $V_{DS}$ and ground, respectively. The bottom edge was fixed at $T_0 = 300$ K and $T_1 = T_2 = 0$ K. The edges along the top were not mechanically constrained. Varying electrical and thermal boundary conditions on the sides from insulating to grounds and heat sinks did not affect predictions.

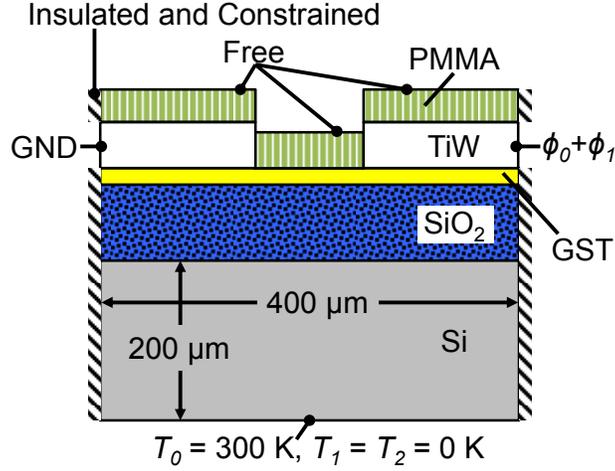

Fig S3. Schematic of model geometry. The model geometry is similar to experiments, with the Si domain modeled as shown. All of the boundaries are electrically and thermally insulated and mechanically constrained unless otherwise noted. The 10 nm SiO$_2$ capping layer was not modeled as it did not affect predictions.

The domain size is ~5 times the Si thermal diffusion length $L_{PD} = [k/(\rho \cdot c_P \cdot f_H)]^{1/2}$. The heating frequency $f_H = f = 28$ kHz for unipolar sine waves and $f_H = 2f = 56$ kHz for bipolar sine waves, as in Fig. 1(b). The material properties of Si are listed in Table SI and $L_{PD} \approx 40$ μm for $f_H = 28$ kHz. Thermo-mechanical expansion of Si accounts for ~32 and ~56 % of the predicted $\Delta h$ for the $L = 1.5$ and 7 μm devices. Thermo-mechanical expansion of PMMA accounts for ~63 and ~41 % of the predicted $\Delta h$ for the $L = 1.5$ and 7 μm devices. Therefore, the GST, TiW, and SiO$_2$ account for ~5 % of the predicted $\Delta h$. We estimate the spatial resolution of SJEM for the GST devices to be similar to our previous work with graphene[S3] due to similar sample geometry and the small contribution of GST to lateral thermal conductance and thermo-mechanical expansion. Decreasing the domain $<L_{PD}$ decreases the temperature rise and thermo-mechanical expansion of the Si, which decreases the predicted $\Delta h$. The device temperature can also decrease due to increased substrate conductance. However, a large thermal resistance between the device and Si substrate can reduce the Si temperature rise and thermo-mechanical expansion to negligible values as shown by our previous work with carbon nanotubes (CNTs).[S4]

Figure S3 shows the modeled geometry. A PMMA thickness of 60 nm was measured on the TiW using an AFM to measure the topography of a scratch through the PMMA on the contact. The TiW contacts were measured to be 80 and 100 nm thick with and without the PMMA by AFM. Therefore, the PMMA thickness was 80 nm in the channel. The PMMA geometry was replicated in the simulation. The measured and modeled PMMA surfaces are similar except within 200 nm of the TiW edge. The measured PMMA has a smooth profile over the TiW edge, instead of the discontinuous profile modeled. The model predicted a dip in $\Delta h$ near the TiW edge, which was not observed in measurements. The discrepancy between predictions and measurements for the dip is attributed to the change in heat flow and thermo-mechanical expansion with the two different PMMA geometries. Therefore, predictions within 200 nm of the edge were ignored.

## 6. Model Properties

Table SI lists the material properties used in the simulation. The electrical resistivity $\rho$ and thermopower $S$ for PMMA, SiO$_2$ (insulator), and Si (substrate) were not included in simulation. The GST-TiW and GST-SiO$_2$ thermal interface conductance $G_{CON} = 4 \times 10^7$ W m$^{-2}$ K$^{-1}$, similar to previous studies.[S5] Varying the GST-TiW $G_{CON} = 10^{7-9}$ W m$^{-2}$ K$^{-1}$ had little effect on predictions. Typical values of $G_{CON}$ for the Si-SiO$_2$, TiW-SiO$_2$, SiO$_2$-PMMA and GST-SiO$_2$ interfaces had little effect on predictions and were neglected. The 10 nm SiO$_2$ capping layer had no effect on predictions and was not modeled. The TiW thermopower was assumed small compared to $S_{GST}$.

Table SI. Electro-Thermo-Physical properties of materials used in the simulation. From left to right the listed properties are the thermal conductivity, density, heat capacity, electrical resistivity, thermopower, coefficient of thermal expansion, Poisson's ration, and elastic modulus.

| Material | $k$ W m$^{-1}$ K$^{-1}$ | $\rho_d$ kg m$^{-3}$ | $c_P$ J kg$^{-1}$ K$^{-1}$ | $\rho$ $\Omega$ m | $S$ $\mu$V K$^{-1}$ | $\alpha_{CTE}$ K$^{-1}$ ×10$^6$ | $v$ | $E$ GPa |
|---|---|---|---|---|---|---|---|---|
| GST | 0.5 [S6] | 6,300 [S6] | 200 [S6] | Measured | Measured | 17 [S7] | 0.3 [S7] | 36 [S7] |
| TiW | 5 | 16,000 | 160 | 1.3×10$^{-5}$ | 1 | 5 [S8] | 0.28 [S8] | 410 [S8] |
| PMMA | 0.18 [S3] | 1,200 [S9] | 1,500 [S9] | - | - | 50 [S4] | 0.35 [S4] | 3 [S4] |
| SiO$_2$ | 1.4 [S3] | 2,220 [S10] | 745 [S10] | - | - | 0.5 [S4] | 0.17 [S4] | 64 [S4] |
| Si | 80 [S11] | 2,330 [S10] | 712 [S10] | - | - | 2.6 [S4] | 0.28 [S4] | 165 [S4] |

Figure S4 shows the effect of varying the material properties on the predicted $\Delta h$ and $\Delta T$ for the $L$ = 1.5 µm device for $V$ = +3.2 V. Decreasing $\rho_C$ = 3×10$^{-10}$ $\Omega$ m$^2$ significantly decreases $\Delta h$ and $\Delta T$ in the contacts due to less power dissipation at the GST-TiW interface. Decreasing the TiW thermal conductivity $k_{TiW}$ = 1 W m$^{-1}$ K$^{-1}$ decreases heat spreading into the contacts. Therefore, $\Delta h$ and $\Delta T$ decrease quickly as $|x|$ increases in the contacts. Increasing $S_{GST}$ = 1 mV K$^{-1}$ increases Peltier heating and cooling of the left and right contacts. The spike in $\Delta h$ and $\Delta T$ at the left contact is due to a hot spot which develops at the GST-TiW interface due to increased heat generation from increased $S_{GST}$ or decreased heat spreading due to decreased $k_{TiW}$. Figure S1(b) shows a higher $\rho_C$ (at the left contact compared to the right contact) will also form a hot spot at the respective GST-TiW interface due to increased heat generation at the left contact. Setting the TiW thermopower $S_{TiW} = S_{GST}$ eliminates Peltier effects in our simulation as there is no difference in thermopower at the GST-TiW interface. We do not observe Thomson heating in our devices, as the negative Thomson coefficient of fcc GST[S12] would heat the GST near the TiW in a manner opposite to our SJEM measurements.

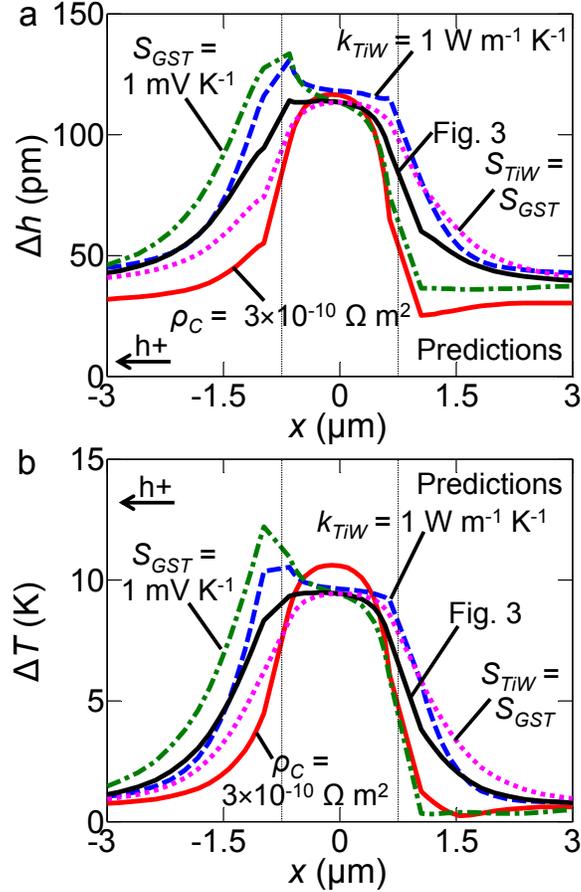

FIG. S4. Effect of model properties on predicted (a) $\Delta h$ and (b) $\Delta T$ for $V = +3.2$ V for the $L = 1.5$ μm device. The solid black line is from Fig. 3 with properties from Table SI. The red solid line, blue dashed line, green dash-dot line, and magenta dotted line vary $\rho_C = 3\times10^{-10}$ Ω m², $S_{GST} = 1$ mV K$^{-1}$, $k_{TiW} = 1$ W m$^{-1}$ K$^{-1}$ and $S_{TiW} = S_{GST}$ from Fig. 3. The black arrow indicates the direction of hole current flow.

## 7. Mean absolute percent error (*MAPE*)

Figure S5 shows the mean absolute percent error (*MAPE*) of the predicted $\Delta h_{Peltier}$ to measurements. The *MAPE* is calculated at both contacts for $V = 6.3$ and $8.9$ V for the $L = 7$ μm device. The predicted $S_{GST} = 450$ and $400$ μV K$^{-1}$ for $V = 6.3$ and $8.9$ V. Figure S5 can be compared to Fig. S2(b) to determine $S_{GST}$. The *MAPE* and $R^2$ are two methods for determining the error in the predicted and measured $\Delta h_{Peltier}$. The *MAPE* for $V = 3.6$ V and the $L = 1.5$ μm device are >100 % and are not shown.

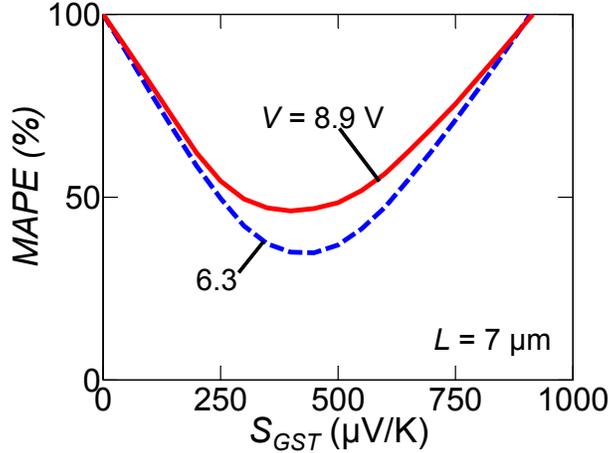

FIG. S5. The mean absolute percent error (*MAPE*) for predicted $\Delta h_{Peltier}$ for $L = 7$ μm device. Shown are $V = 6.3$ and 8.9 V in blue dashed and red solid lines.

## 8. TiW Properties

Van der Pauw measurements were used to evaluate the sputtered TiW film (1 cm$^2$ sample, 100 nm thickness, 10/90 % wt, annealed at 200 $^o$C similar to GST devices) and found an electrical resistivity $\rho_{TiW} = 1.3 \times 10^{-5}$ Ω m. Using this value, the Wiedemann-Franz law predicts an estimated electron thermal conductivity $k_{el} \approx 0.6$ W m$^{-1}$ K$^{-1}$ at room temperature. The phonon thermal conductivity $k_{ph}$ can be estimated from kinetic theory as:

$$k_{ph} = 1/3 C_{ph} v \lambda \qquad (SE7)$$

Where $C_{ph}$, $v$, and $\lambda$ are the phonon heat capacity, phonon velocity, and phonon mean free path. The Debye model is used in Eqn. SE8 to approximate $C_{ph}$.

$$C_{ph} = 9 n_a k_B \left(\frac{T}{T_D}\right)^3 \int_0^{T_D/T} \frac{x^4 e^x dx}{(e^x - 1)^2} \qquad (SE8)$$

Where $n_a$, $k_B$ and $T$ are the atom number density, Boltzmann constant, and temperature. Assuming a lattice constant $a \approx 3.1$ Å, approximately the lattice spacing of Ti and W, and a bcc struc-ture for TiW[S13] we calculate $n_a \sim 4 \times 10^{24}$ kg$^{-1}$ and the density $\rho_d$ along with the heat capacity are listed in Table SI. The Debye temperature $T_D \sim 400$ K is found as:

$$T_D = \frac{\hbar v}{k_B} \left( 6\pi^2 n_a \right)^{1/3} \tag{SE9}$$

where $\hbar$ is the reduced Planck's constant and the speed of sound in W is used to estimate $v \approx 3.2 \times 10^3$ m s$^{-1}$. Varying $v = 2\text{-}5 \times 10^3$ m s$^{-1}$ and $a = 2.8\text{-}3.2$ Å adjusts $k_{ph}$ by at most 50%. The largest uncertainty of $k_{ph}$ is due to the estimation of the phonon mean free path $\lambda$. Based on the relatively high $\rho_{TiW}$ of our films, we estimate $\lambda \sim 1\text{-}2$ nm, as there is likely a large amount of disorder in the TiW which scatters both electrons and phonons. X-ray diffraction (XRD) measurements indeed suggest the grain size is ~3 nm. Finally, the combined TiW thermal conductivity is $k_{TiW} = k_{el} + k_{ph} \approx 5$ W m$^{-1}$ K$^{-1}$. Figures 3(a) and S1(a) show that predictions based on $k_{TiW} \approx 5$ W m$^{-1}$ K$^{-1}$ agree well with our SJEM measurements.

## 9. GST Phase

Figure S6 shows X-ray diffraction (XRD) measurements of the GST films after the 5 min 200 °C anneal. The peaks present at the (111), (200), and (220) labels indicate the film is in the fcc state. Slight shifts in the XRD peak locations are due to strain or small changes in stoichiometry. The relative intensities of the fcc peaks change due to preferred direction of growth. The calculated ~8 nm grain size is smaller than 15-20 nm in previous work.[S6] Smaller grains yield shallow and broad peaks in the XRD measurements compared to previous work.[S6,S14,S15] The low peak intensity can indicate the presence of the amorphous phase. However, the measured resistivity and anneal conditions are similar to previous work indicating the film is predominantly fcc.[S14,S15]

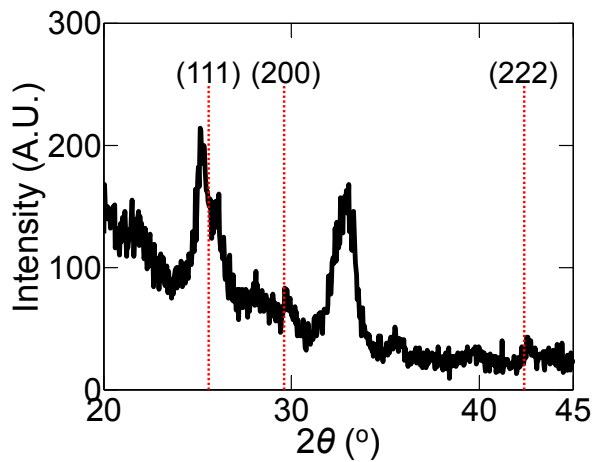

FIG. S6. X-ray diffraction (XRD) measurements of the 25-nm thick GST film after the 5 min. anneal at 200 °C. Red dashed vertical lines are the predicted peak locations for fcc GST. The peak at ~33° is from the underlying Si.

## Supplemental References